\documentclass[
    ,final            
  ]
  {aipproc}

\usepackage{times,amssymb,mathptmx}

\newcommand{\simgt}{\lower.5ex\hbox{$\; \buildrel > \over \sim \;$}}
\newcommand{\simlt}{\lower.5ex\hbox{$\; \buildrel < \over \sim \;$}}

\layoutstyle{8x11single}

\begin{document}

\title{Recent progress on understanding ``pasta'' phases in dense stars}


\author{Gentaro Watanabe}{
  address={NORDITA, Blegdamsvej 17, DK-2100 Copenhagen \O, Denmark}
  ,altaddress={The Institute of Chemical and Physical Research (RIKEN),
Saitama 351-0198, Japan}
}

\author{Hidetaka Sonoda}{
  address={Department of Physics, University of Tokyo,
Tokyo 113-0033, Japan}
}


\begin{abstract}
In cores of supernovae and crusts of neutron stars,
nuclei can adopt interesting shapes, such as rods or slabs, etc.,
which are referred to as nuclear ``pasta.''
Recently, we have been studying the pasta phases focusing
on their dynamical aspects with quantum molecular dynamic (QMD) approach.
We review our findings on the following topics: 
dynamical formation of the pasta phases 
by cooling down the hot uniform nuclear matter;
a phase diagram on the density versus temperature plane;
structural transitions between the pasta phases 
induced by compression and their mechanism.
Properties of the nuclear interaction used in our works are also discussed.
\end{abstract}

\maketitle


\section{I.\quad Introduction}

In ordinary matter, atomic nuclei are roughly spherical.
This may be understood in the liquid drop picture of the nucleus as being a 
result of the forces due to the surface tension of nuclear matter,
which favors a spherical nucleus, being greater than those due to the 
electrical repulsion between protons, which tends to make the nucleus deform.
When the density of matter approaches that of atomic nuclei, i.e.,
the normal nuclear density $\rho_0$, nuclei are closely packed
and the effect of the electrostatic energy becomes comparable
to that of the surface energy.
Consequently, at subnuclear densities around $\rho\simeq\rho_0/2$,
the energetically favorable configuration is expected to have
remarkable structures:
the nuclear matter region (i.e., the liquid phase)
is divided into periodically arranged parts of
rodlike or slablike shape, embedded in the gas phase and
in a roughly uniform electron gas.
Besides, there can be phases in which nuclei are turned inside out,
with cylindrical or spherical bubbles of the gas phase
in the liquid phase.
These phases with nonspherical nuclei
are often referred to as nuclear ``pasta'' phases
because nuclear slabs and rods look like ``lasagna'' and ``spaghetti.''
Likewise, spherical nuclei and bubbles are called
``meatballs'' and ``cheese,'' respectively.

In equilibrium dense matter in supernova cores and neutron stars,
existence of the pasta phases has been predicted by 
Ravenhall {\it et al.} \cite{rpw} and Hashimoto {\it et al.} \cite{hashimoto}.
Since these seminal works, properties of the pasta phases 
in equilibrium states have been investigated with various nuclear models.
They include studies on phase diagrams at zero temperature
\cite{lorenz,oyamatsu,sumiyoshi,gentaro,williams}
and at finite temperatures \cite{lassaut}.
These earlier works have confirmed that, for various nuclear models,
the nuclear shape changes as:
  $\mbox{sphere} \rightarrow \mbox{cylinder} \rightarrow \mbox{slab}
  \rightarrow \mbox{cylindrical hole} \rightarrow \mbox{spherical hole}
  \rightarrow \mbox{uniform}$,
with increasing density.

In these earlier works, however,
a liquid drop model or the Thomas-Fermi approximation
is used with an assumption on the nuclear shape
(except for Ref.\ \cite{williams}).
Thus the phase diagram at subnuclear densities
and the existence of the pasta phases should be examined
without assuming the nuclear shape.
It is also noted that
at temperatures of several MeV, which are relevant to the collapsing cores,
effects of thermal fluctuations
on the nucleon distribution are significant.
However, these thermal fluctuations cannot be described properly by
mean-field theories such as the Thomas-Fermi approximation
used in the previous work \cite{lassaut}.

In contrast to the equilibrium properties, dynamical or non-equilibrium aspects
of the pasta phases had not been studied until recently 
except for some limited cases \cite{formation,review}.
Thus it had been unclear even whether or not the pasta phases can be formed and
the transitions between them can be realized 
during the collapse of stars and the
cooling of neutron stars, which have finite time scales.

To solve the above problems, molecular dynamic approaches
for nucleon many-body systems are suitable.
They treat the motion of the nucleonic degrees of freedom
and can describe thermal fluctuations and many-body correlations
beyond the mean-field level.

Using the framework of QMD \cite{aichelin}, 
which is one of the molecular dynamic methods, 
we have solved the following two major questions
\cite{qmd_transition,qmd_cold,qmd_hot}.
\begin{itemize}
\item {\it Question} 1: 
Whether or not the pasta phases are formed by cooling down hot
uniform nuclear matter in a finite time scale much smaller than
that of the neutron star cooling?
\item {\it Question} 2: 
Whether or not transitions between the pasta phases can occur
by the compression during the collapse of a star?
\end{itemize}

The pasta phases have recently begun to attract the
attention of many researchers 
(see, e.g., Refs.\ \cite{burrows,martinez} and references therein).
The mechanism of the collapse-driven supernova
explosion has been a central mystery in astrophysics for
almost half a century (e.g., Ref.\ \cite{bethe}). 
Previous studies suggest that the
revival of the shock wave by neutrino heating is a crucial
process. As has been pointed out in Refs.\ \cite{gentaro,qmd_cold} 
and elaborated in Refs.\ \cite{horowitz1,horowitz2,future}, 
the existence of the pasta
phases instead of uniform nuclear matter increases the
neutrino opacity of matter in the inner core significantly
due to the neutrino coherent scattering by nuclei \cite{freedman,sato};
this affects the total energy transferred to the
shocked matter. Thus the pasta phases could play an important
role in the future study of supernova explosions.
Our recent work \cite{qmd_transition}
strongly suggests the possibility of dynamical formation
of the pasta phases from a crystalline lattice of spherical nuclei;
effects of the pasta phases on the supernova explosions
should be seriously discussed in the near future.

\section{II.\quad Method: Quantum Molecular Dynamics}

Among various versions of the molecular dynamic models,
Quantum Molecular Dynamics (QMD) \cite{aichelin} is the most practical one
for investigating the pasta phases.
Rodlike and slablike nuclei are mesoscopic entities of nuclei themselves
and they contain a large number of nucleons.
QMD, which is a less elaborate in the treatment of the exchange effect,
allows us to study such large systems with several nonspherical nuclei.
The typical length scale $r_{\rm c}$ of half of the inter-structure
is $r_{\rm c} \sim 10$ fm and the density region of interest is
around half of the normal nuclear density $\rho_{0}$.
The total nucleon number $N$ necessary to reproduce $n$ structures
in the simulation box is $N \sim \rho_{0} (2r_{\rm c}n)^{3}$ (for slabs).
It is thus desirable to use $\sim 10^4$ nucleons
in order to reduce boundary effects.
Such large systems are difficult to be handled by other molecular dynamic
models such as FMD \cite{fmd} and AMD \cite{amd},
whose calculation costs increase as $\sim N^4$,
but are tractable for QMD, whose calculation costs increase as $\sim N^2$.

It is also noted that the exchange effect is less important
for the nuclear pasta structures, 
which are in the macroscopic scale for nucleons.
This can be seen by comparing the typical values of
the exchange energy
and of the energy difference between pasta phases.
Suppose there are two identical nucleons, $i=1$ and 2, bound in
different nuclei.
The exchange energy between these particles is calculated
as an exchange integral:
$I= \int U({\bf r}_{1}-{\bf r}_{2})\
\varphi_{1}({\bf r}_{1}) \varphi_{1}^{*}({\bf r}_{2})
\varphi_{2}({\bf r}_{2}) \varphi_{2}^{*}({\bf r}_{1})\
d{\bf r}_{1} d{\bf r}_{2}\ $,
where $U$ is the potential energy.
An asymptotic form of the wave function is given by
$\varphi_{i} \sim \exp{(- k_{i} r)}$ 
with 
$k_{i} = \sqrt{2mE_{i}}/\hbar\ ,\ ( i=1,2 )$,
where $E_{i}$ is the binding energy and $m$ is the nucleon mass.
The exchange integral reads
$I \sim \exp{[-(k_{1}+k_{2})R]} \sim 5 \times 10^{-6}$ MeV
for the internuclear distance $R \simeq 10$ fm
and $E_{i} \simeq 8$ MeV,
which is extremely smaller than the typical energy difference
per nucleon between different pasta phases
of order 0.1 keV 
(for neutron star matter) - 10 keV (for supernova matter).
Therefore, it is expected that QMD
is not a bad approximation for investigating the pasta phases.

\subsection{1.\quad Model Hamiltonian and its Properties}

In our studies on the pasta phases, we have used a nuclear force
given by a QMD model Hamiltonian with the medium-equation-of-state
parameter set in Ref.\ \cite{maruyama}.
This model Hamiltonian consists of six parts:
\begin{equation}
  {\cal H} =
  K+V_{\rm Pauli}+V_{\rm Skyrme}+V_{\rm sym}+V_{\rm MD}+V_{\rm Coulomb}\ ,
  \label{hamiltonian}
\end{equation}
where $K$ is the kinetic energy;
$V_{\rm Pauli}$ is the momentum-dependent ``Pauli potential,'' 
which reproduces the effects of the Pauli principle phenomenologically;
$V_{\rm Skyrme}$ is the Skyrme potential
which consists of an attractive two-body term and a repulsive three-body term;
$V_{\rm sym}$ is the symmetry potential;
$V_{\rm MD}$ is the momentum-dependent potential
introduced as two Fock terms of the Yukawa interaction;
$V_{\rm Coulomb}$ is the Coulomb energy between protons.

The parameters in the Pauli potential are determined to fit the kinetic energy
of the free Fermi gas at zero temperature.
The above model Hamiltonian reproduces the binding energy of 
symmetric nuclear matter,
16 MeV per nucleon, at the normal nuclear density $\rho_0=0.165$ fm$^{-3}$
and other saturation properties: the incompressibility is set to be
280 MeV and the symmetry energy is 34.6 MeV.
This model Hamiltonian also well reproduce the properties of stable nuclei
relevant to our interest:
the binding energy except for light nuclei from $^{12}$C to $^{20}$Ne 
\cite{maruyama},
and the rms radius of the ground state of heavy ones with $A \gtrsim 100$
\cite{kido}.
It is also confirmed that another QMD Hamiltonian close to this model
provides a good description of nuclear reactions including the low energy
region (several MeV per nucleon) \cite{niita}.

Let us then examine other properties of the nuclear interaction
(at zero temperature),
which have not been determined accurately yet but
have important effects on inhomogeneous structure of matter
at subnuclear densities.
Such quantities are the nuclear surface tension $E_{\rm surf}$,
the energy per nucleon $\epsilon_{\rm n}$ of
the pure neutron matter, and the proton chemical potential $\mu_{\rm p}^{(0)}$
in the pure neutron matter.
The surface tension $E_{\rm surf}$, which
is the most important among these three quantities,
controls the size of the nuclei and bubbles,
and hence the sum of the Coulomb and surface energies.
With increasing $E_{\rm surf}$ and so this energy sum,
the density $\rho_{\rm m}$ at which matter becomes uniform is lowered.
There is a tendency, especially in a case of neutron star matter,
that the higher $\epsilon_{\rm n}$, $\rho_{\rm m}$ is lowered.
This is because larger $\epsilon_{\rm n}$ tends to favor
uniform nuclear matter without dripped neutron gas regions than mixed phases.
In neutron star matter, there is also a tendency that
the lower $\mu_{\rm p}^{(0)}$, the smaller $\rho_{\rm m}$.
This is because $-\mu_{\rm p}^{(0)}$ represents the
degree to which the neutron gas outside the nuclei
favors the presence of protons in itself.

\begin{figure}[htbp]
\rotatebox{0}{
\resizebox{8.2cm}{!}
{\includegraphics{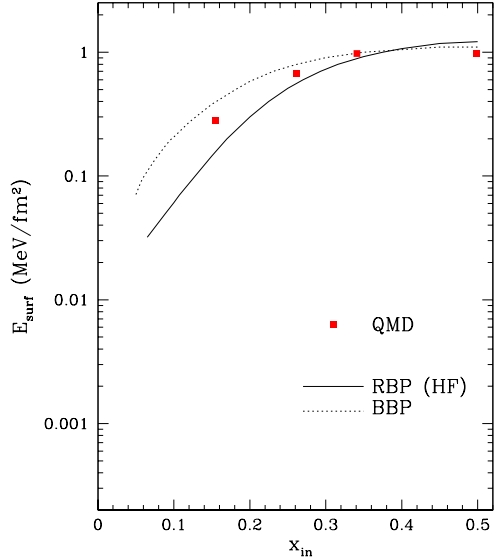}}}
\caption{\label{fig esurf}(Color)\quad
  The nuclear surface energy
  per unit area (the surface tension) $E_{\rm surf}$ versus
  the proton fraction $x_{\rm in}$ in the nuclear matter region.
  The red solid squares are the values of the present QMD Hamiltonian
  \cite{maruyama};
  the solid curve is the result of the Skyrme-Hartree-Fock calculation
  with a modified version of 1' parameter set
  done by Ravenhall, Bennett and Pethick (RBP) \cite{rbp};
  the dotted curve is from Baym, Bethe and Pethick (BBP) \cite{bbp}.
  This figure is adapted from Ref.\ \cite{qmd_cold}.
  }
\end{figure}

\begin{figure}[htbp]
\rotatebox{0}{
\resizebox{16.4cm}{!}
{\includegraphics{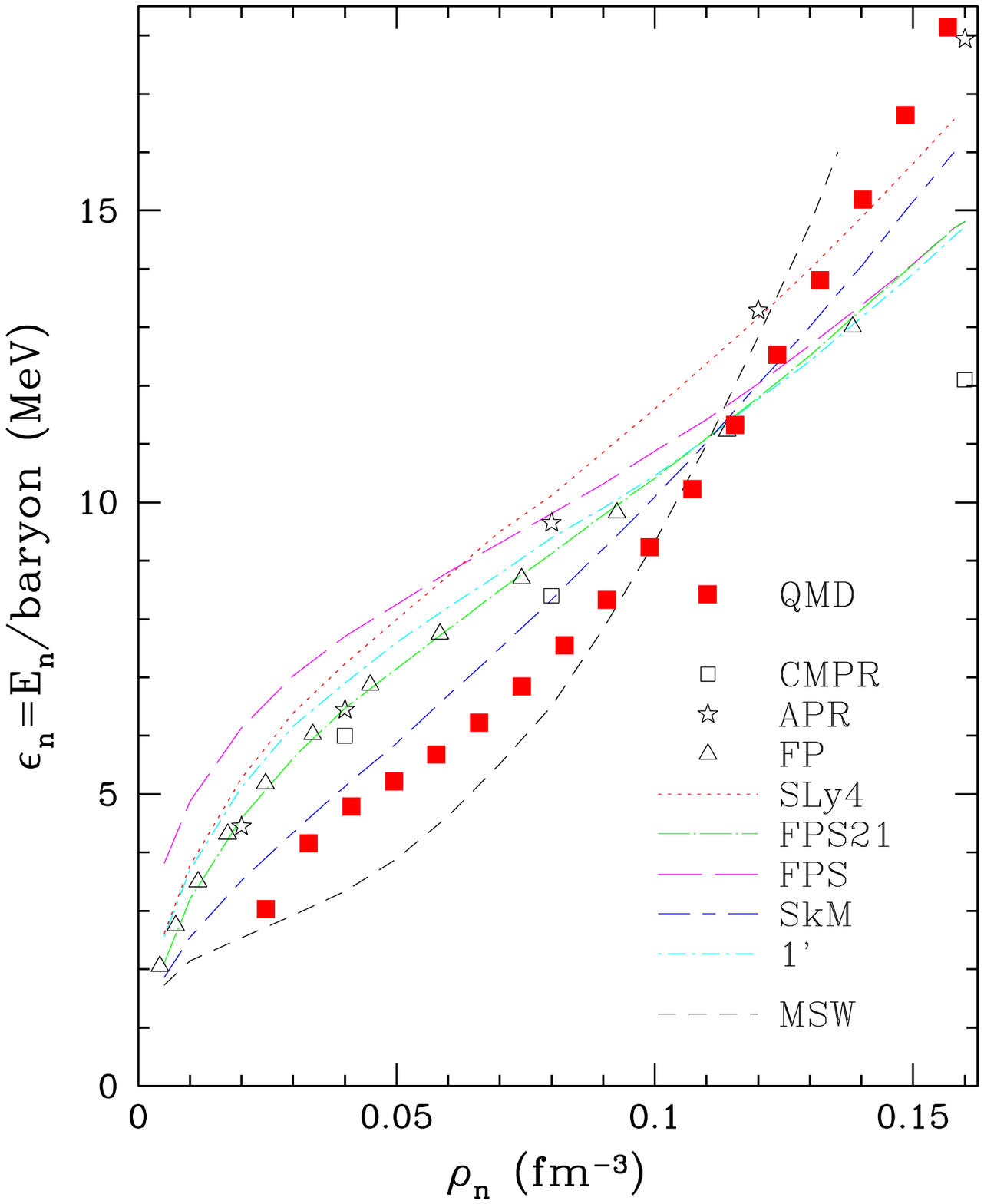}\includegraphics{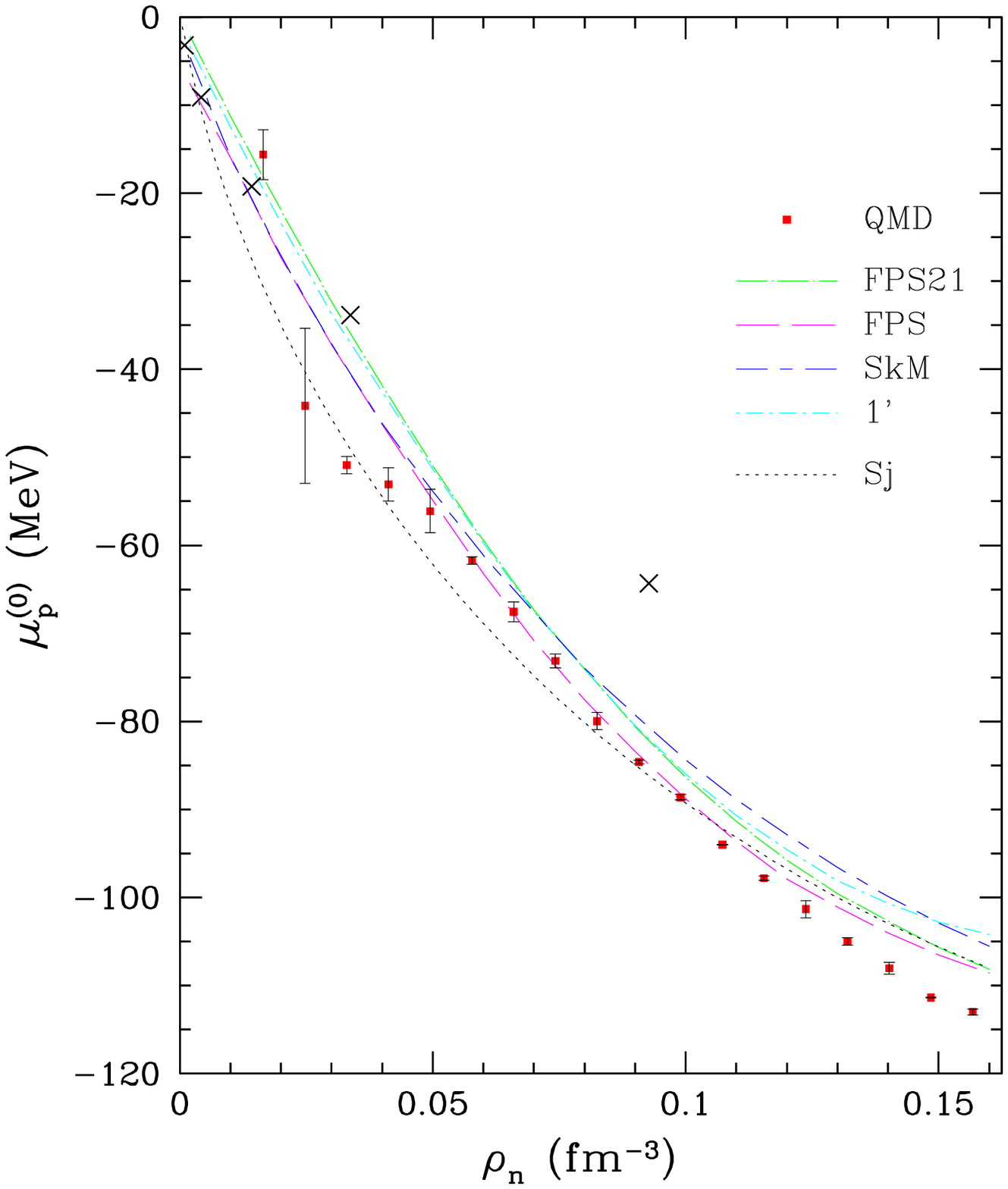}}}%
\caption{\label{fig nmatter}(Color)\quad
  The neutron density $\rho_{\rm n}$ dependence of
  the energy per nucleon $\epsilon_{\rm n}$ (left panel) and 
  the proton chemical potential $\mu_{\rm p}^{(0)}$ 
  (right panel) of pure neutron matter.
  The red solid squares show the result of the present QMD model Hamiltonian
  \cite{maruyama}.
  The red dotted line denoted by SLy4 is the result from Ref.\ \cite{douchin},
  and the colored broken lines as marked by the other Skyrme interactions
  (FPS21, 1', FPS and SkM) are the results summarized by
  Pethick, Ravenhall and Lorenz \cite{prl}.
  The black dotted line is the result of Sj\"oberg \cite{sj} and
  the black dashed line is that of Myers, Swiatecki and Wang \cite{msw}.
  The open squares shows the result of the GFMC calculation 
  by Carlson {\it et al.} \cite{gfmc};
  the open stars denote the values obtained by 
  Akmal, Pandharipande and Ravenhall \cite{akmal};
  the triangles are those from Friedman and Pandharipande \cite{fp};
  the crosses from Siemens and Pandharipande \cite{siemens}.
  The large error bars and the scatter of $\mu_{\rm p}^{(0)}$ of QMD
  in the low density region $\rho \lesssim 0.3 \rho_0$
  are due to the local roughness of the density in the neutron matter, 
  which is caused by the fixed width of the wave packet.
  This figure is adapted from Ref.\ \cite{qmd_cold}.
  }
\end{figure}

In Figs.\ \ref{fig esurf} and \ref{fig nmatter},
we have plotted the results of these quantities 
for the present model Hamiltonian.
We can say that, on the whole, they give reasonable values
within uncertainties of each quantities.
It is noted that $E_{\rm surf}$ of the present model shows
moderate values between the RBP and BBP results.
The behavior of $\mu_{\rm p}^{(0)}$ is also in a reasonable agreement
with various Skyrme-Hartree-Fock calculations except for 
higher densities of $\rho_{\rm n}\gtrsim 0.1$ fm$^{-3}$
relevant only for neutron star matter just below $\rho_{\rm m}$.
The quantity $\epsilon_{\rm n}$, however, shows
some deviation from the major trend of Skyrme-Hartree-Fock results
and of the values of the microscopic calculations.
The behavior of $\epsilon_{\rm n}$ of the present model 
is similar to that of the SkM interaction and 
of the interaction by Myers {\it et al.} \cite{msw}.

In the following, for each case of different value 
of the proton fraction $x$ of matter,
we summarize the consequences of the features of 
$E_{\rm surf}$, $\epsilon_{n}$ and $\mu_{p}^{(0)}$
of the present QMD interaction.
\begin{itemize}
  \item[1.] For symmetric nuclear matter ($x=x_{\rm in}=0.5$)

    According to $E_{\rm surf}$ at $x_{\rm in}=0.5$,
    the present model is consistent with the other results,
    and is an appropriate effective interaction
    for studying the pasta phases at $x=0.5$.

  \item[2.] For neutron star matter ($x \lesssim 0.1$)

    The melting density $\rho_{\rm m}$ is lowered by larger $E_{\rm surf}$,
    steep rise of $\epsilon_{\rm n}$ at $\rho\gtrsim 0.12$ fm$^{-3}$ and
    larger negative values of $\mu_{\rm p}^{(0)}$ 
    at $\rho\gtrsim 0.1$ fm$^{-3}$ compared to various Skyrme-Hartree-Fock
    calculations.

  \item[3.] For supernova matter ($x \sim 0.3$)

    Relatively low $\epsilon_{\rm n}$ 
    at low neutron densities $\lesssim 0.1$ fm$^{-3}$
    acts to favor mixed phases rather than the uniform phase
    and $E_{\rm surf}$ acts in the opposite way
    in comparison with the Skyrme-Hartree-Fock result by RBP \cite{rbp}.
\end{itemize}

\section{III.\quad Simulations and Results}

Using the framework of QMD, we have solved the two major questions
posed in the beginning of this article \cite{qmd_transition,qmd_cold,qmd_hot}.
In the present section, we will review these works
\footnote{This section is based on our recent review article
\cite{soft review}.}.
Hereafter, we set the Boltzmann constant $k_{\rm B}=1$.

In our simulations,
we treated the system which consists of neutrons, protons, and electrons
in a cubic box with periodic boundary condition.
The system is not magnetically polarized,
i.e., it contains equal numbers of protons (and neutrons) with spin up and
spin down.
The relativistic degenerate electrons which ensure charge neutrality
are regarded as a uniform background \cite{review}
(see Refs.\ \cite{screening,screen_maruyama} 
for effects of the electron screening).
The Coulomb interaction is calculated by the Ewald method 
taking account of the Gaussian charge distribution of the proton wave packets.

\subsection{1.\quad Realization of the Pasta Phases
and Equilibrium Phase Diagrams\label{subsect_realization}}

In Refs.\ \cite{qmd_cold,qmd_hot}, we have reproduced the pasta phases
from hot uniform nuclear matter and discussed phase diagrams
at zero and finite temperatures.
In these works, we first prepared a uniform hot nucleon gas
at the temperature $T \sim 20$ MeV as an initial condition,
which is equilibrated for $\sim 500 - 2000$ fm/$c$ in advance.
To realize the ground state of matter,
we then cooled it down slowly until the temperature got $\sim 0.1$ MeV
or less for $O(10^{3}-10^{4})$ fm/$c$,
keeping the nucleon number density constant.
In the cooling process, we mainly used the frictional relaxation method
(equivalent to the steepest descent method), which is given by
the QMD equations of motion plus small friction terms.
In the case of finite temperatures, we also used
thermostat to reproduce the equilibrium states.

\begin{figure}[t]
\resizebox{12cm}{!}{\includegraphics{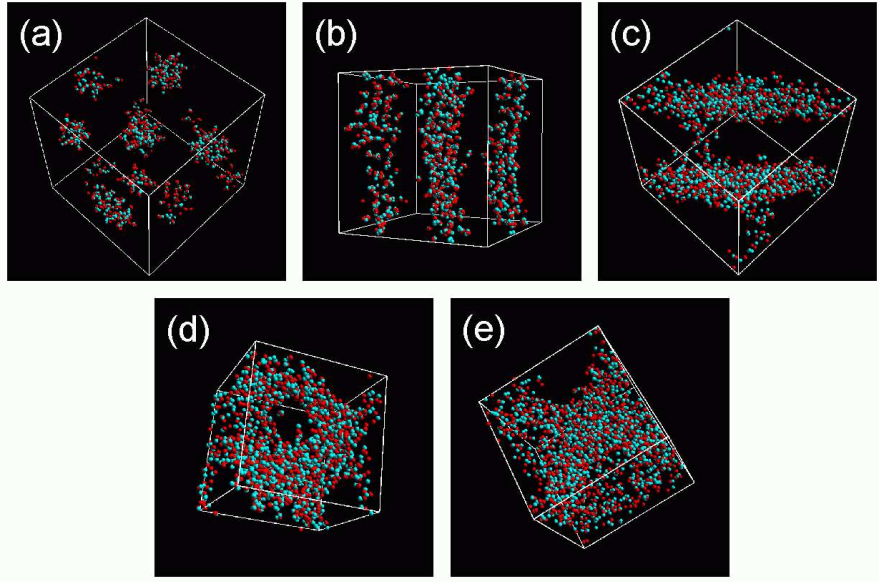}}
\caption{\label{fig pasta sym}(Color)\quad
  Nucleon distributions of the pasta phases in cold matter at $x=0.5$;
  (a) sphere phase, $0.1 \rho_{0}$ ($L_{\rm box}=43.65\ {\rm fm}$, $N=1372$);
  (b) cylinder phase, $0.225 \rho_{0}$ ($L_{\rm box}=38.07\ {\rm fm}$, $N=2048$);
  (c) slab phase, $0.4 \rho_{0}$ ($L_{\rm box}=31.42\ {\rm fm}$, $N=2048$);
  (d) cylindrical hole phase, $0.5 \rho_{0}$ ($L_{\rm box}=29.17\ {\rm fm}$, $N=2048$) and
  (e) spherical hole phase, $0.6 \rho_{0}$ ($L_{\rm box}=27.45\ {\rm fm}$, $N=2048$),
  where $L_{\rm box}$ is the box size and $N$ is the total number of nucleons.
  The whole simulation box is shown in this figure.
  The red particles represent protons and the green ones neutrons.
  Taken from Ref.\ \cite{qmd_hot}.
  }
\end{figure}

The resultant typical nucleon distributions of cold matter 
at subnuclear densities are shown
in Fig.\ \ref{fig pasta sym}
for proton fraction of matter $x=0.5$.
We can see from these figures that
the phases with rodlike and slablike nuclei,
cylindrical and spherical bubbles,
in addition to the phase with spherical nuclei are reproduced.
The above simulations have shown that the pasta phases 
can be formed dynamically from hot uniform matter
within a time scale of $\sim O(10^{3}-10^{4})$ fm/$c$.

\begin{figure}[htbp]
\resizebox{13cm}{!}
{\includegraphics{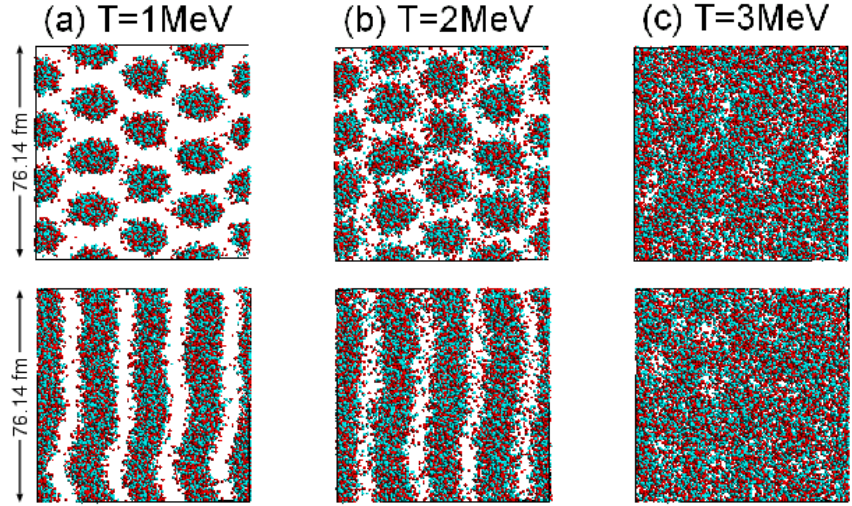}}
\caption{\label{snap 0.225rho x0.5 16000}(Color)\quad
  Nucleon distributions for $x=0.5$, $\rho=0.225\rho_{0}$
  at the temperatures of 1, 2 and 3MeV.
  The total number of nucleons $N=16384$ 
  and the box size $L_{\rm box}=76.14$ fm.
  The upper panels show the top views along the axis of 
  the cylindrical nuclei at $T=0$, the lower ones the side views.
  Protons are represented by the red particles, and
  neutrons by the green ones.
  Taken from Ref.\ \cite{qmd_hot}.
  }
\end{figure}

We show snapshots of nucleon distributions at $T=1, 2$ and 3 MeV
for a density $\rho=0.225\rho_0$ in Fig.\ \ref{snap 0.225rho x0.5 16000}.
This density corresponds to the phase with rodlike nuclei at $T=0$.
From these figures, we can see the following qualitative features:
at $T\simeq 1.5-2$ MeV
(but snapshots for $T\simeq 1.5$ MeV are not shown),
the number of the evaporated nucleons becomes significant; 
at $T\gtrsim 3$ MeV, 
nuclei almost melt and the spatial distribution of the nucleons
are rather smoothed out.

\begin{figure}[htbp]
\rotatebox{270}{
\resizebox{7.5cm}{!}
{\includegraphics{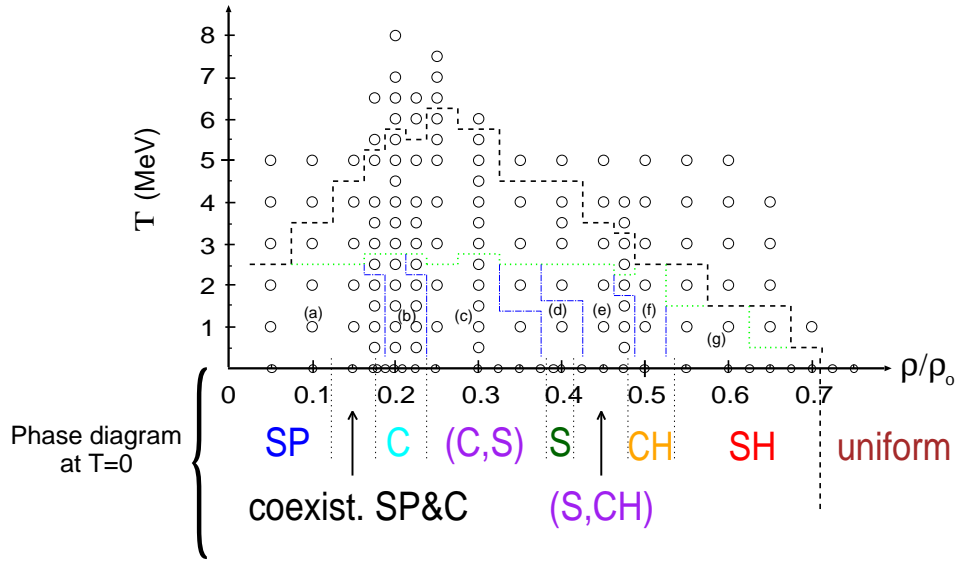}}}
\caption{\label{phase diagram x0.5}(Color)\quad
  Phase diagram of matter at $x=0.5$ plotted in the $\rho$ - $T$ plane.
  The dashed and the dotted lines on the diagram
  show the phase separation line and
  the limit below which the nuclear surface can be identified, respectively.
  The dash-dotted lines are the phase boundaries between
  the different nuclear shapes.
  The symbols SP, C, S, CH, SH, U stand for nuclear shapes,
  i.e., sphere, cylinder, slab, cylindrical hole,
  spherical hole and uniform, respectively.
  The parentheses (A,B) show intermediate phases between A and B-phases
  with negative $\chi$, which are different from coexistence phases of them.
  The regions (a)-(g) correspond to the nuclear shapes characterized by
  $\int_{\partial R} H dA $ and $\chi$ as follows:
  (a) $\int_{\partial R} H dA > 0,\ \chi > 0$;
  (b) $\int_{\partial R} H dA > 0,\ \chi = 0$;
  (c) $\int_{\partial R} H dA > 0,\ \chi < 0$;
  (d) $\int_{\partial R} H dA = 0,\ \chi = 0$;
  (e) $\int_{\partial R} H dA < 0,\ \chi < 0$;
  (f) $\int_{\partial R} H dA < 0,\ \chi = 0$;
  (g) $\int_{\partial R} H dA < 0,\ \chi > 0$.
  Simulations have been carried out at points denoted by circles.
  Adapted from Ref.\ \cite{qmd_cold}
  }
\end{figure}

When we try to classify the nuclear structure systematically,
the integral mean curvature and the Euler characteristic
(see, e.g., Ref.\ \cite{minkowski} and references therein)
are useful. 
Suppose there is a set of regions $R$,
where the density is higher than a threshold density $\rho_{\rm th}$. 
The integral mean curvature and the Euler characteristic
for the surface of this region $\partial R$
are defined as surface integrals of
the mean curvature $H = (\kappa_{1}+\kappa_{2})/2$ and
the Gaussian curvature $G = \kappa_{1} \kappa_{2}$, respectively;
i.e., $\int_{\partial R} H dA$ and
$\chi \equiv \frac{1}{2 \pi} \int_{\partial R} G dA$,
where $\kappa_{1}$ and $\kappa_{2}$ are the principal curvatures and
$dA$ is the area element of the surface of $R$.
The Euler characteristic $\chi$ depends only on the topology of $R$
and is expressed as
$\chi =$ (number of isolated regions)
$-$ (number of tunnels) $+$ (number of cavities).
Using a combination of these two quantities calculated for nuclear surface
\footnote{Nuclear surface generally corresponds to an isodensity surface
for the threshold density $\rho_{\rm th}\simeq 0.5\rho_0$ 
in our simulations.}, 
each pasta phase can be represented uniquely, i.e.,
for the phase with
spherical nuclei:  $\int_{\partial R} H dA  > 0,\ \chi > 0$,
cylindrical nuclei:  $\int_{\partial R} H dA  > 0,\ \chi = 0$,
slablike nuclei:  $\int_{\partial R} H dA  = 0,\ \chi = 0$,
cylindrical bubbles:  $\int_{\partial R} H dA  < 0,\ \chi = 0$,
and spherical bubbles:  $\int_{\partial R} H dA  < 0,\ \chi > 0$.
We note that the value of $\chi$ for the ideal pasta phases
is zero except for the phase with spherical nuclei or spherical bubbles
with positive $\chi$; 
negative $\chi$ is not obtained for the pasta phases.

The phase diagram obtained for $x=0.5$ is plotted in 
Fig.\ \ref{phase diagram x0.5}.
As shown above,
nuclear surface can be identified
typically at $T \lesssim 3$ MeV (see the dotted lines)
in the density range of interest.
Thus the regions between the dotted line and the dashed line
correspond to some non-uniform phase,
which is however difficult to be classified
into specific phases because the nuclear surface cannot be identified well.

In the region below the dotted lines,
where we can identify the nuclear surface, we have obtained 
the pasta phases with spherical nuclei [region (a)], 
rodlike nuclei [region (b)], slablike nuclei [region (d)],
cylindrical holes [region (f)] and spherical holes [region (g)].
It is noted that in addition to these pasta phases,
structures with negative $\chi$ have been also obtained
in the regions of (c) and (e); matter consists of 
multiply connected nuclear and bubble regions (i.e., spongelike structure) with
branching rodlike nuclei, perforated slabs and branching bubbles, etc.
A detailed discussion on the phase diagrams is given in Ref.\ \cite{qmd_hot}.

\subsection{2.\quad Structural Transitions between the Pasta Phases
\label{subsect_transition}}

In Ref.\ \cite{qmd_transition}, we have approached the second question
asked at the beginning of this article.
We have performed QMD simulations of the compression of dense matter
and have succeeded in simulating the transitions between 
rodlike and slablike nuclei and between slablike 
nuclei and cylindrical bubbles.

The initial conditions of the simulations are samples of the columnar phase
($\rho=0.225 \rho_0$) and of the laminar phase ($\rho=0.4 \rho_0$)
of 16384-nucleon system at $x=0.5$ and $T\simeq1$ MeV.
These are obtained in Ref.\ \cite{qmd_hot},
which are presented in the last section.
We then adiabatically compressed the above samples by increasing the density
at the average rate of $\simeq$1.3$\times 10^{-5} \rho_0/$(fm$/c$)
for the initial condition of the columnar phase and
$\simeq$7.1$\times 10^{-6} \rho_0/$(fm$/c$) for that of the laminar one.
According to the typical value of the density difference
between each pasta phase, $\sim 0.1\rho_0$ 
(see Fig.\ \ref{phase diagram x0.5}),
we increased the density to the value corresponding to the next pasta phase
taking the order of $10^4$ fm$/c$, which was much longer than
the typical time scale of the nuclear fission, $\sim 1000$ fm$/c$.
Thus the above rates ensured the adiabaticity 
of the simulated compression process
with respect to the change of nuclear structure.
Finally, we relaxed the compressed sample
at $\rho=0.405 \rho_0$ for the former case and at $0.490 \rho_0$ 
for the latter one.
These final densities are those
of the phase with slablike nuclei and cylindrical bubbles, respectively,
in the equilibrium phase diagram at $T \simeq 1$ MeV 
(see Fig.\ \ref{phase diagram x0.5}).

\begin{figure}[t]
\resizebox{13cm}{!}
{\includegraphics{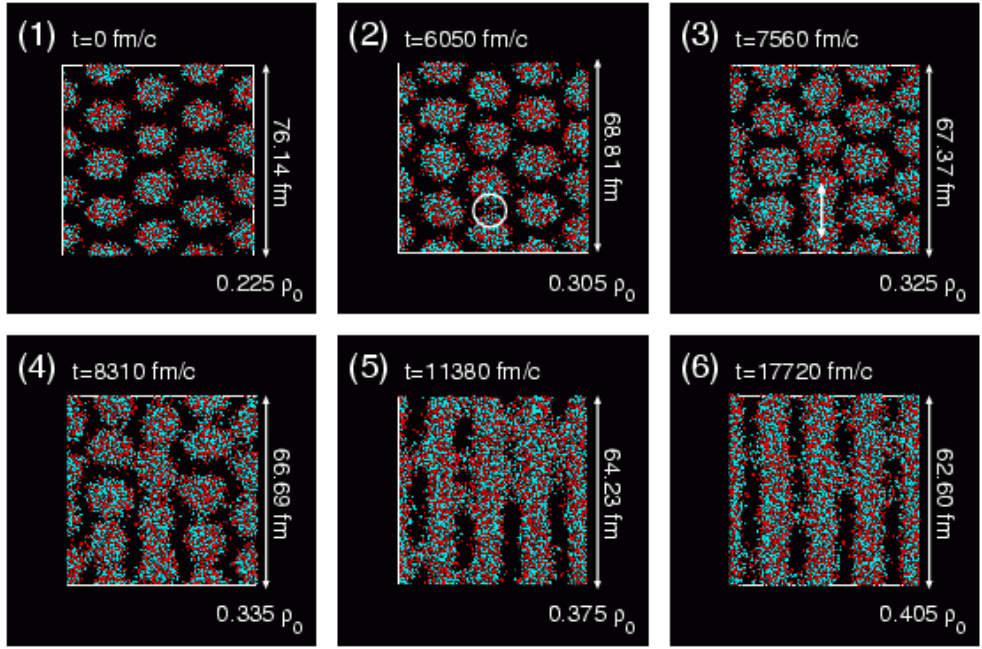}}
\caption{\label{fig rod slab}(Color)\quad 
Snapshots of the transition process from
the phase with rodlike nuclei to the phase with slablike nuclei
(the whole simulation box is shown).
The red particles show protons and the green ones neutrons.
After neighboring nuclei touch as shown by the circle in 
Fig.\ \ref{fig rod slab}-(2),
the ``compound nucleus'' elongates along the arrow in 
Fig.\ \ref{fig rod slab}-(3).
The box size is rescaled to be equal in this figure.
Adapted from Ref.\ \cite{qmd_transition}.
}
\end{figure}

\begin{figure}[t]
\resizebox{13cm}{!}
{\includegraphics{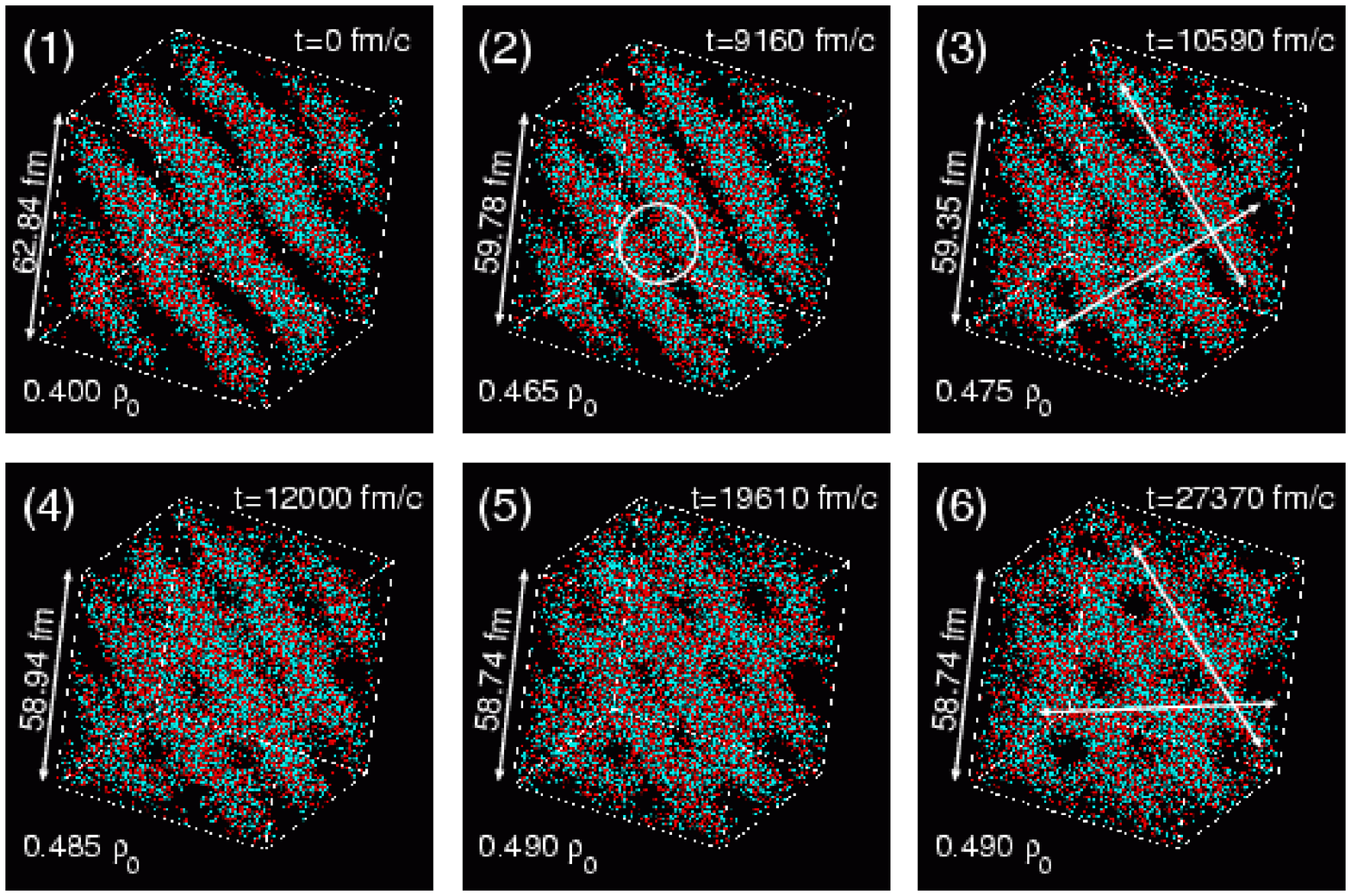}}
\caption{\label{fig slab cylindhole}(Color)\quad 
The same as Fig.\ 1 for the transition from
the phase with slablike nuclei to the phase with cylindrical holes
(the box size is not rescaled in this figure).
After the slablike nuclei begin to touch [see the circle in 
Fig.\ \ref{fig slab cylindhole}-(2)],
the bridges first crosses them almost orthogonally as shown by the arrows
in Fig.\ \ref{fig slab cylindhole}-(3). 
Then the cylindrical holes are formed and they
relax into a triangular lattice, as shown by the arrows in 
Fig.\ \ref{fig slab cylindhole}-(6).
Adapted from Ref.\ \cite{qmd_transition}.
}
\end{figure}

The resulting time evolution of the nucleon distribution
is shown in Figs.\ \ref{fig rod slab} and \ref{fig slab cylindhole}.
As can be seen from Fig.\ \ref{fig rod slab},
the phase with slablike nuclei is finally formed 
[Fig.\ \ref{fig rod slab}-(6)]
from the phase with rodlike nuclei [Fig.\ \ref{fig rod slab}-(1)].
The temperature in the final state is $\simeq 1.35$ MeV.
It is noted that the transition is triggered by thermal fluctuation,
not by the fission instability:
when the internuclear spacing becomes small enough 
and once some pair of neighboring rodlike nuclei 
touch due to thermal fluctuations,
they fuse [Figs.\ \ref{fig rod slab}-(2) and \ref{fig rod slab}-(3)].
Then such connected pairs of rodlike nuclei further touch and fuse with 
neighboring nuclei in the same lattice plane like a chain reaction
[Fig.\ \ref{fig rod slab}-(4)]; the time scale of the each fusion process
is of order $10^2$ fm$/c$, which is much smaller than 
that of the density change.

The transition from the phase with slablike nuclei
to the phase with cylindrical holes is shown in 
Fig.\ \ref{fig slab cylindhole}.
When the internuclear spacing decreases enough,
neighboring slablike nuclei touch due to the thermal fluctuation
as in the above case.
Once nuclei begin to touch [Fig.\ \ref{fig slab cylindhole}-(2)]
bridges between the slabs are formed at many places
on a time scale (of order $10^2$ fm$/c$)
much shorter than that of the compression.
After that the bridges cross the slabs
nearly orthogonally for a while [Fig.\ \ref{fig slab cylindhole}-(3)].
Nucleons in the slabs continuously flow into the bridges,
which become wider and merge together to form cylindrical holes.
Afterwards, the connecting regions
consisting of the merged bridges move gradually,
and the cylindrical holes relax to form a triangular lattice 
[Fig.\ \ref{fig slab cylindhole}-(6)].
The final temperature in this case is $\simeq 1.3$ MeV.

\begin{figure}[t]
\rotatebox{270}{
\resizebox{!}{13cm}
{\includegraphics{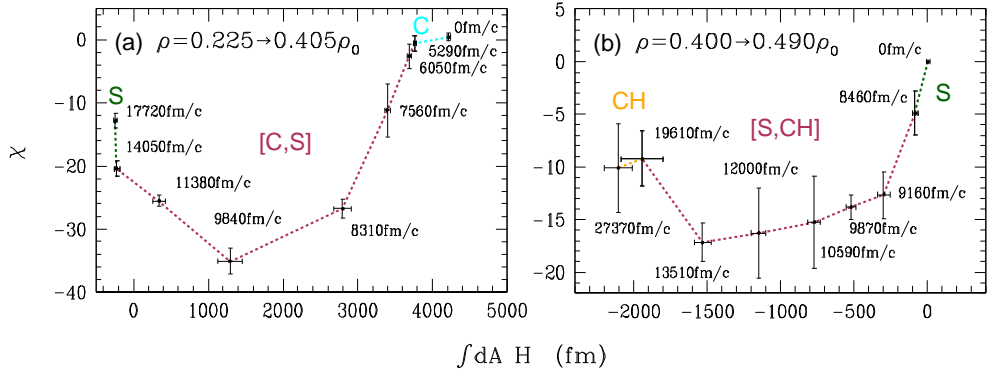}}}
\caption{\label{fig curv euler}(Color)\quad 
Time evolution of $\int_{\partial R} H dA$ and $\chi$
during the simulations.
The data points and the error bars show, respectively, the mean values
and the standard deviations in the range of the threshold density
$\rho_{\rm th}=0.3$ -- $0.5 \rho_0$,
which includes typical values for the nuclear surface.
The panel (a) is for the transition from cylindrical (C)
to slablike nuclei (S)
and the panel (b) for the transition from slablike nuclei
to cylindrical holes (CH).
Transient states are shown as [C,S] and [S,CH] for each transition.
Adapted from Ref.\ \cite{qmd_transition}.
}
\end{figure}

Trajectories of the above processes on the plane of 
the integral mean curvature $\int_{\partial R} H dA$ 
and the Euler characteristic $\chi$ are plotted in Fig.\ \ref{fig curv euler}.
This figure shows that the above transitions proceed through a transient
state with ``spongelike'' structure, which gives negative $\chi$.
As can be seen from Fig.\ \ref{fig curv euler}-(a) 
[Fig.\ \ref{fig curv euler}-(b)],
the value of the Euler characteristic begins to decrease from zero
when the rodlike [slablike] nuclei touch. It continues to decrease
until all of the rodlike [slablike] nuclei are connected to others
by small bridges at $t\simeq 9840$ fm$/c$ [$\simeq 12000$ fm$/c$].
Then the bridges merge to form slablike nuclei [cylindrical holes]
and the value of the Euler characteristic increases towards zero.
Finally, the system relaxes into a layered lattice
of the slablike nuclei [a triangular lattice of the cylindrical holes].
Thus the whole transition process can be divided into
the ``connecting stage'' and the ``relaxation stage''
before and after the moment at which the Euler characteristic is minimum; 
the former starts when the nuclei begin to touch and it
takes $\simeq 3000$ -- 4000 fm$/c$ and the latter takes
more than 8000 fm$/c$.

\subsection{3.\quad Formation of the Pasta Phases
\label{subsect_formation}}

\begin{figure}[t]
\resizebox{9cm}{!}
{\includegraphics{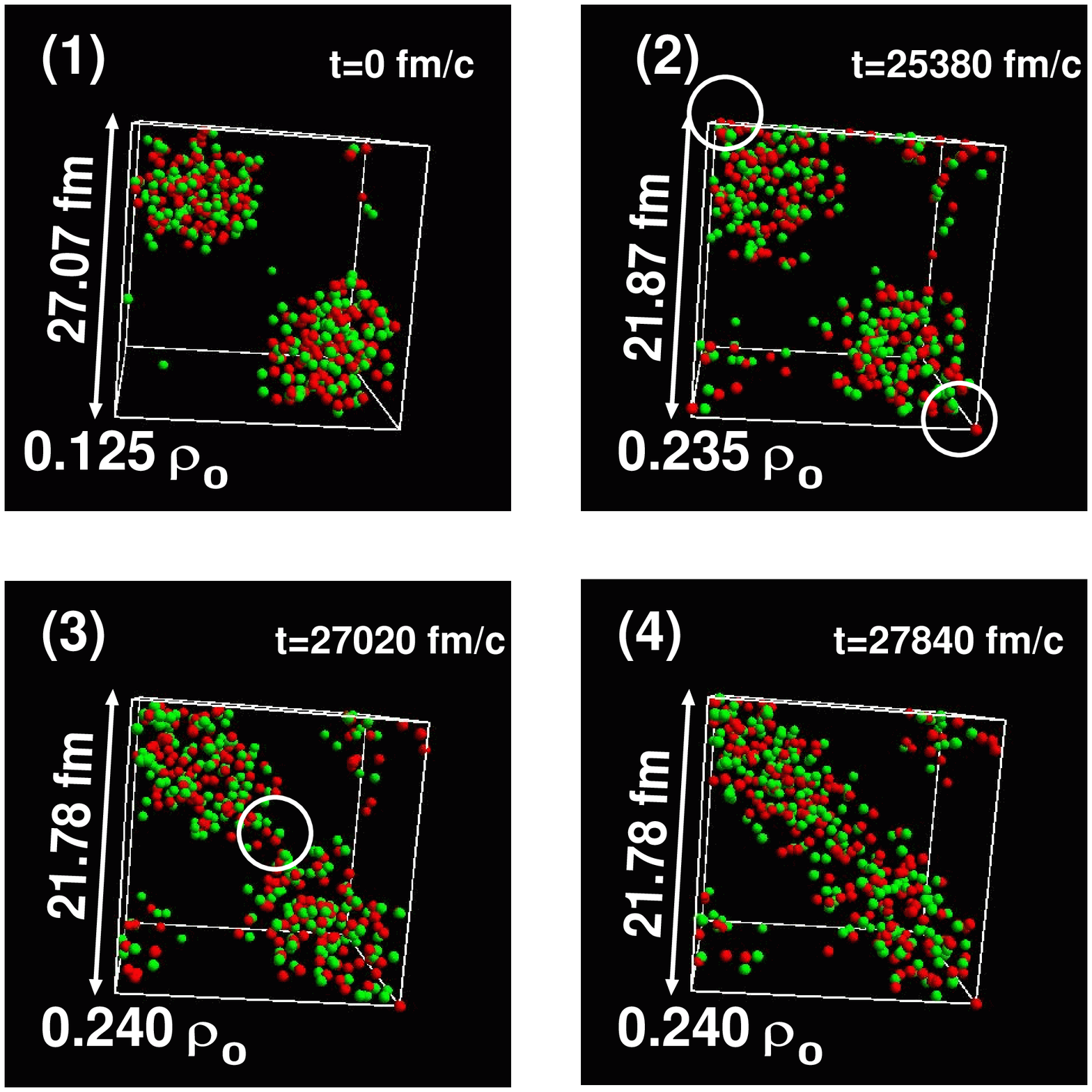}}
\caption{\label{fig sphere rod}(Color)\quad 
Snapshots of the transition process from
the bcc lattice of spherical nuclei to the triangular lattice of rodlike nuclei
(the whole simulation box is shown).
The red particles show protons and the green ones neutrons.
The box size is rescaled to be equal in this figure.
}
\end{figure}

In closing the present article, let us briefly show our recent results
of a study on the formation process of pasta nuclei
from spherical ones; i.e., a transition from the phase with spherical nuclei
to that with rodlike nuclei.
Time evolution of the nucleon distribution in the transition process
is shown in Fig.\ \ref{fig sphere rod}.
The initial condition of this simulation is 
a nearly perfect bcc unit cell with 409 nucleons (202 protons and
207 neutrons) at $T\simeq 1$ MeV.
We compressed the system in a similar way to that of the simulations
explained in the previous section.
The average rate of the density change in the present case is
$\simeq$4.4$\times 10^{-6} \rho_0/$(fm$/c$).
Since the two nuclei start to touch [see the circles in 
Fig.\ \ref{fig sphere rod}-(2)], the transition process completes
within $\simeq 2500$ fm$/c$ and the rodlike nucleus is formed.
The final state [Fig.\ \ref{fig sphere rod}-(4)]
is a triangular lattice of the rodlike nuclei.

The present simulation has been performed using a rather small system;
effects of the finite system size in this simulation should be examined.
Detailed investigation of the transition using a larger system
will be presented in a future publication \cite{future}.

\section{IV.\quad Conclusion}

We approached the two questions posed in Section I using the framework of QMD.
According to the results of our simulations, our answer is 
strongly affirmative for both questions.
The nuclear interaction used in our simulations
shows generally reasonable properties at subnuclear densities
not only for symmetric nuclear matter but also for neutron matter.
This result also supports our conclusion.






\begin{theacknowledgments}
G. W. appreciates C. J. Pethick for his valuable comments and
hospitality at NORDITA.
The research reported in this article grew out of collaborations with
Kei Iida, Toshiki Maruyama, Katsuhiko Sato, Kenji Yasuoka and
Toshikazu Ebisuzaki.
Further research currently in progress is performed
using RIKEN Super Combined Cluster System with MDGRAPE-2.
This work was supported in part
by the Nishina Memorial Foundation,
by the JSPS Postdoctoral Fellowship for Research Abroad,
by the Japan Society for the Promotion of Science,
by the Ministry of
Education, Culture, Sports, Science and Technology
through Research Grant No. 14-7939,
and by RIKEN through Research Grant No. J130026.
\end{theacknowledgments}


\begin{thebibliography}{99}
%
\bibitem{aichelin} J. Aichelin and H. St{\"o}cker,
  Phys.\ Lett. {\bf B176}, 14 (1986); J. Aichelin,
  Phys.\ Rep. {\bf 202}, 233 (1991).
%
\bibitem{akmal} A. Akmal, V. R. Pandharipande and D. G. Ravenhall,
  Phys.\ Rev.\ C {\bf 58}, 1804 (1998).
%
\bibitem{bbp} G. Baym, H. A. Bethe and C. J. Pethick,
  Nucl.\ Phys. {\bf A175}, 225 (1971).
%
\bibitem{bethe} H. A. Bethe,
  Rev.\ Mod.\ Phys. {\bf 62}, 801 (1990).
%
\bibitem{burrows} A. Burrows, S. Reddy and T. A. Thompson,
  Nucl.\ Phys. {\bf A}, in press (astro-ph/0404432).
%
\bibitem{gfmc} J. Carlson, J. Morales, Jr., V. R. Pandharipande and D. G. Ravenhall
  Phys.\ Rev.\ C {\bf 68}, 025802 (2003).
%
\bibitem{douchin} F. Douchin and P. Haensel, 
  Phys.\ Lett. {\bf B485}, 107 (2000).
%
\bibitem{fmd} H. Feldmeier,
  Nucl.\ Phys. {\bf A515}, 147 (1990); H. Feldmeier and J. Schnack,
  Prog.\ Part.\ Nucl.\ Phys. {\bf 39}, 393 (1997).
%
\bibitem{freedman} D. Z. Freedman,
  Phys.\ Rev.\ D {\bf 9}, 1389 (1974).
%
\bibitem{fp} B. Friedman and V. R. Pandharipande,
  Nucl.\ Phys. {\bf A361}, 502 (1981).
%
\bibitem{hashimoto} M. Hashimoto, H. Seki and M. Yamada,
  Prog.\ Theor.\ Phys. {\bf 71}, 320 (1984).
%
\bibitem{horowitz1} C. J. Horowitz, M. A. P\'erez-Garc\'ia, and J. Piekarewicz,
  Phys.\ Rev.\ C {\bf 69}, 045804 (2004).
%
\bibitem{horowitz2} C. J. Horowitz, M. A. P\'erez-Garc\'ia, J. Carriere, D. K. Berry, and J. Piekarewicz,
  Phys.\ Rev.\ C {\bf 70}, 065806 (2004).
%
\bibitem{formation} K. Iida, G. Watanabe and K. Sato,
  Prog.\ Theor.\ Phys. {\bf 106}, 551 (2001);
  Erratum, {\it ibid.} {\bf 110}, 847 (2003).
%
\bibitem{kido} T. Kido, T. Maruyama, K. Niita and S. Chiba,
  Nucl.\ Phys. {\bf A663 \& 664}, 877c (2000).
%
\bibitem{lassaut} M. Lassaut, H. Flocard, P. Bonche, P.H. Heenen and E. Suraud,
  Astron.\ Astrophys. {\bf 183}, L3 (1987).
%
\bibitem{lorenz} C. P. Lorenz, D. G. Ravenhall and C. J. Pethick,
  Phys.\ Rev.\ Lett. {\bf 70}, 379 (1993).
%
\bibitem{martinez} G. Martinez-Pinedo, M. Liebendoerfer, D. Frekers,
  astro-ph/0412091.
%
\bibitem{maruyama} T. Maruyama, K. Niita, K. Oyamatsu, T. Maruyama,
  S. Chiba and A. Iwamoto,
  Phys.\ Rev.\ C {\bf 57}, 655 (1998).
%
\bibitem{screen_maruyama} T. Maruyama, T. Tatsumi, D. N. Voskresensky, T. Tanigawa, and S. Chiba,
  Phys.\ Rev.\ C {\bf 72}, 015802 (2005).
%
\bibitem{minkowski} K. Michielsen and H. De Raedt,
  Phys.\ Rep. {\bf 347}, 461 (2001).
%
\bibitem{msw} W. D. Myers, W. J. Swiatecki and C. S. Wang,
  Nucl.\ Phys. {\bf A436}, 185 (1985).
%
\bibitem{niita} K. Niita,
  in the Proceedings of the Third Simposium on
  {\it ``Simulation of Hadronic Many-body System''},
  A. Iwamoto {\it et al.}, Eds.,
  JAERI-conf. {\bf 96-009}, 22 (1996) (in Japanese).
%
\bibitem{amd} A. Ono, H. Horiuchi, T. Maruyama and A. Ohnishi,
  Prog.\ Theor.\ Phys. {\bf 87}, 1185 (1992);
  Phys.\ Rev.\ Lett. {\bf 68}, 2898 (1992).
%
\bibitem{oyamatsu} K. Oyamatsu,
  Nucl.\ Phys. {\bf A561}, 431 (1993).
%
\bibitem{review} C. J. Pethick and D. G. Ravenhall,
  Annu.\ Rev.\ Nucl.\ Part.\ Sci. {\bf 45}, 429 (1995).
%
\bibitem{prl} C. J. Pethick, D. G. Ravenhall and C. P. Lorenz,
  Nucl.\ Phys. {\bf A584}, 675 (1995).
%
\bibitem{rbp} D. G. Ravenhall, C. D. Bennett and C. J. Pethick,
  Phys.\ Rev.\ Lett. {\bf 28}, 978 (1972).
%
\bibitem{rpw} D. G. Ravenhall, C. J. Pethick and J. R. Wilson,
  Phys.\ Rev.\ Lett. {\bf 50}, 2066 (1983).
%
\bibitem{sato} K. Sato,
  Prog.\ Theor.\ Phys. {\bf 53}, 595 (1975);
  {\it ibid.} {\bf 54}, 1325 (1975).
%
\bibitem{siemens} P. J. Siemens and V. R. Pandharipande,
  Nucl.\ Phys. {\bf A173}, 561 (1971).
%
\bibitem{sj} O. Sj\"oberg, 
  Nucl.\ Phys. {\bf A222}, 161 (1974).
%
\bibitem{sumiyoshi} K. Sumiyoshi, K. Oyamatsu and H. Toki,
  Nucl.\ Phys. {\bf A595}, 327 (1995).
%
\bibitem{future} G. Watanabe {\it et al.},
  to be published.
%
\bibitem{screening} G. Watanabe and K. Iida,
  Phys.\ Rev.\ C {\bf 68}, 045801 (2003).
%
\bibitem{gentaro} G. Watanabe, K. Iida and K. Sato,
  Nucl.\ Phys. {\bf A676}, 455 (2000);
  {\it ibid.} {\bf A687}, 512 (2001);
  Erratum, {\it ibid.} {\bf A726}, 357 (2003).
%
\bibitem{qmd_transition} G. Watanabe, T. Maruyama, K. Sato, K. Yasuoka
  and T. Ebisuzaki,
  Phys.\ Rev.\ Lett. {\bf 94}, 031101 (2005).
%
\bibitem{qmd_cold} G. Watanabe, K. Sato, K. Yasuoka and T. Ebisuzaki,
  Phys.\ Rev.\ C {\bf 66}, 012801(R) (2002); 
  {\it ibid.} {\bf 68}, 035806 (2003).
%
\bibitem{qmd_hot} G. Watanabe, K. Sato, K. Yasuoka and T. Ebisuzaki,
  Phys.\ Rev.\ C {\bf 69}, 055805 (2004).
%
\bibitem{soft review} G. Watanabe and H. Sonoda,
  to appear in ``Soft Condnsed Matter: New Research'',
  ed. F. Columbus (cond-mat/0502515).
%
\bibitem{williams} R. D. Williams and S. E. Koonin,
  Nucl.\ Phys. {\bf A435}, 844 (1985).
%
\end{thebibliography}
\end{document}